# Study of systematic effects in beta decay measurements with $AgReO_4$ calorimeters

Alessandro Monfardini,[a*] Francesca Capozzi,[b] Oliviero Cremonesi,[b] Angelo Nucciotti,[b] Monica Sisti,[b]

[a]*INFN gruppo di Trento and ITC-IRST, via Sommarive 18, 38050 Povo (TN), Italy*

[b]*INFN sez. Milano and Università di Milano-Bicocca,Dipartimento di Fisica, piazza Della Scienza 3, 20126 Milano, Italy*

**Abstract**

$AgReO_4$ microcalorimeters are planned to be used again in the next generation of calorimetric neutrino mass experiments with sensitivity down to 0.2 eV. The understanding and characterization of all possible sources of systematic uncertainties is crucial. In this work we focus on two of these sources, which have been studied in the 10 detectors of the Milano $AgReO_4$ array experiment (MIBETA): a) the solid-state Beta Environmental Fine Structure (BEFS) observed for the first time in $AgReO_4$; b) the detector energy response for internal beta events, which has been investigated with a dedicated measurement using a $^{44}$Ti gamma source. The possible effects on neutrino mass experiments due to the incomplete understanding of these two aspects are discussed. 

*Keywords*: microbolometers, neutrino mass, beta decay, cryogenic detectors.
*PACS*: 23.40, 14.60.Pq

## 1. Introduction

The discovery of neutrino flavour oscillations is probably the most convincing piece of evidence for physics beyond the Standard Model [1]. Besides the intrinsic importance, those observations also boosted the interest for a direct measure of the neutrino mass.

The classical, kinematical method making use of the distortions in the electronic beta spectrum still seems the best way to approach the problem. So far, the best upper limit is $2.2 eV/c^2$ [2] obtained with an electrostatic spectrometer. Major systematic effects affecting spectrometer experiments are related to the separation between the source and the detector. An alternative approach has then to ensure at least that source ≡ detector. An elegant way out is to make them both to coincide with the absorber of an ultra-cryogenic (*T*<100mK) bolometer [3]. Every decaying nucleus determines here a transient, measurable $\Delta T$. The most stringent upper limit determined with this technique relies on the lowest-*Q* beta decay [4]:

$$^{187}\mathrm{Re} \rightarrow {}^{187}\mathrm{Os} + e^- + \bar{\nu}_e \quad \left(Q = 2466\,\mathrm{eV}; T_{1/2} = 43.2\,\mathrm{Gyr}\right)$$

* Corresponding author. Tel. +39 0461 314258 Fax +39 0461 314340. e-mail: monfardini@itc.it

In particular, in [4] the source/detector is made up by ten $AgReO_4$ micro-crystals ($m$ = 200÷300μg). More ambitious bolometric experiments have been proposed to explore the eV and sub-eV $m_\nu$ range [5]. In the present paper we describe two potential systematic effects that could affect a high statistics $AgReO_4$ bolometric measure:

- BEFS (Beta Environmental Fine Structure) solid-state effects affecting the spectrum shape;
- detector internal (bulk) response.

## 2. Beta Environmental Fine Structure (BEFS)

In 1991 Koonin [6] suggested, in analogy with EXAFS (Extended X-ray Absorption Fine Structure), that the β electron should have suffered reflections from the surrounding atoms. More precisely, the surrounding atoms influence the decay rate in an energy-dependent (oscillatory) way. In [7] the BEFS theory has been extended to include forbidden beta decays like $^{187}$Re. The first BEFS experimental evidence has been achieved with a bolometric measurement and metallic rhenium absorbers [8]. We present in fig. 1 the measured BEFS effect in $AgReO_4$ [9] together with a suitable model fit extrapolated until the end-point. The lattice zero-point energy and the diminished scattering cross-section both determine a damping of the oscillations superimposed to the "free atom" beta spectrum. For $E_{e^-}$ > 2keV the extrapolated BEFS signal is an oscillation of about $2\cdot10^{-4}$ amplitude with features FWHM of the order of 40eV. The "quasi-period" is determined by the distance of the Re and Ag first atomic shells (4Å and 3.8 Å respectively).

Considering the proposed experiment MARE (Microcalorimeter Arrays for a Rhenium Experiment) [5] phase I ($T_{END} \leq$ 2010), the integrated number of beta decays is around $10^{10}$ to get $m_\nu$ < 2eV. A Montecarlo code has been adopted to confirm that the BEFS effect, even if ignored, is not influencing the neutrino mass sensitivity as far as the fit is performed at energies exceeding 1 keV.

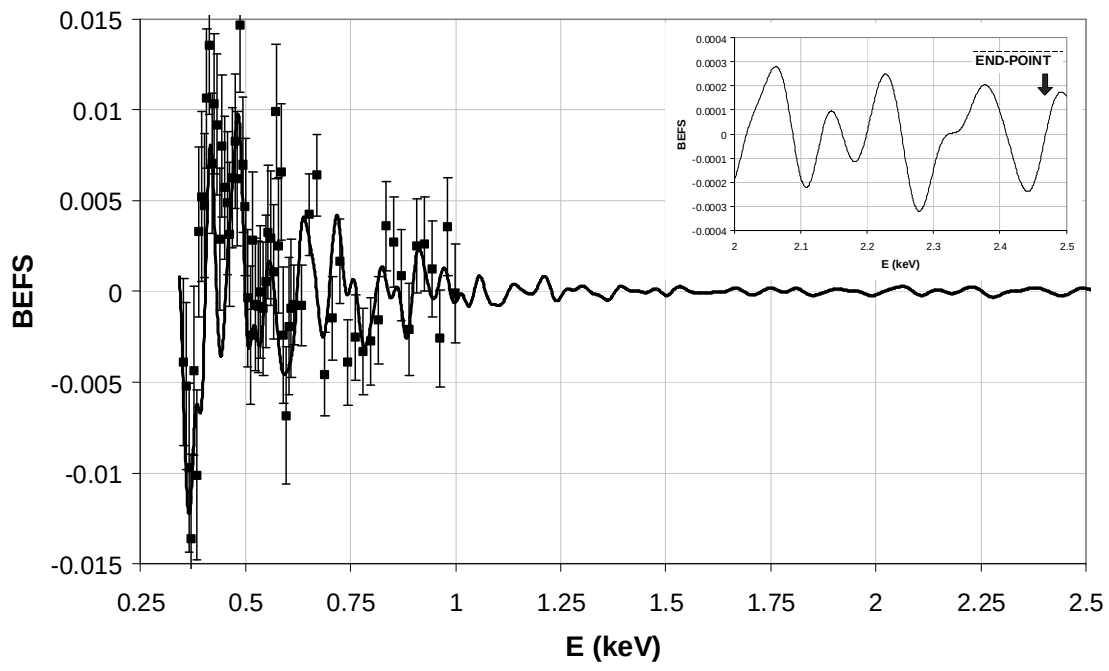

Figure 1. BEFS oscillations observed in $AgReO_4$ (MIBETA). Insert: end-point region (E > 2keV) extrapolation ($\Delta E_{FWHM}$=23eV).

In the MARE phase II ($10^{14}$ events, $m_\nu$ sensitivity ~0.2eV 90% C.L., $T_{END} \geq$ 2015), on the other hand, the BEFS "noise" in the end-point region will be comparable to the poissonian fluctuations. However, the introduction of a corrective mask on the expected spectrum shape seems quite safe taking into account that the effect won't be largely predominant over the natural statistical uncertainties.

It is important to point out that in case of compact/dense absorbers (e.g. metallic rhenium) the BEFS effect amplitude could be increased by more than one order of magnitude. $AgReO_4$, despite the lower active nuclei fraction, seems a relatively safer absorber for sub-eV experiments as far as solid-state effects are considered.

## 3. Uniform internal calibration

In order to provide a realistic estimate of the neutrino mass (or its upper limit), the detector intrinsic response must be precisely determined and understood. An external $^{55}$Fe X-ray calibration source has been adopted so far to deposit a fixed amount of energy within the absorber. The typical range of a 6keV x-ray photon in $AgReO_4$ crystals is about 4µm, to be compared with the typical absorbers dimensions (≈300µm). The energy is thus deposited on the exposed crystal surface with possible border effects (e.g. electron escape). An exponential low-energy tail has in fact been observed on high-statistics calibration lines. A defective understanding of the tail nature would introduce an unacceptable systematic error even for MARE phase I [5].

### 3.1. $^{44}$Ti and Re escape peaks

A $^{44}$Ti radioactive source produces penetrating 78.4keV γ-ray photons able to determine K and L emissions from Re atoms distributed all over the micro-crystal. The escape peaks have energies of 9.05keV ($K_{\beta1}$), 9.34keV ($K_{\beta3}$), 17.22keV ($K_{\alpha2}$) and 18.64keV ($K_{\alpha1}$). The large intrinsic width (42.1eV) represents the main limitation of this approach.

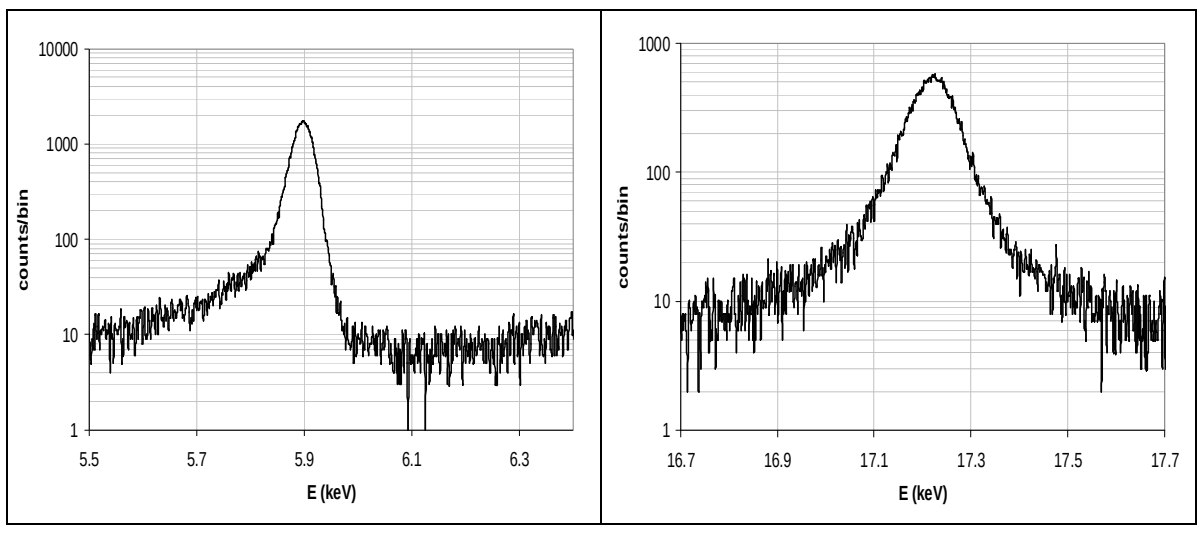

Figure 2. Mn $K_\alpha$ "external" (left) and $^{44}$Ti-induced "internal" Re $K_{\alpha2}$ line (right) from the same calibration run.

A complete analysis is in progress. Preliminary results (fig. 2) seem to confirm that the low-energy tail evident on surface events is less pronounced or lacking on $K_\alpha$ escape peaks.

### 3.2. BEFS as an internal calibration tool

The BEFS peaks positions are fixed and determined uniquely by the $AgReO_4$ local structure. They could represent a powerful self-calibrating tool for energies $E$ < 2keV, where the statistics will allow a precise fitting and a safe modeling.

The FWHM width of the BEFS features is increasing with the electron energy (<10eV @400eV; ≈40eV @2keV). The peaks shapes are peculiar and dependent upon the local structure of $AgReO_4$. Refer to fig. 1 for a theoretical prediction in case of detector energy resolution $\Delta E_{FWHM}$ = 23eV.

## 4. Conclusions

This study demonstrates that the solid-state BEFS effect should not significantly affect future neutrino mass determinations based on $AgReO_4$ calorimeters. For dense/compact absorbers the BEFS oscillations could, on the other hand, be a limiting factor.

The internal detector response cannot be easily determined by adopting an external x-ray source due to the insufficient penetration in the crystal bulk. The preliminary results of a test run adopting an external $^{44}$Ti γ source have been presented. They seem to indicate that the "bulk" response is more symmetric and tail-free. The main limitation of this approach, in case of optimized detectors ($\Delta E_{FWHM}$ ~ 5eV), is the large natural width of the Re escape lines.

An alternative internal calibration, suitable for low-energy ($E$ < 2keV), could make use of the BEFS modulation peaks that are going to be fully resolved and modeled.

## Acknowledgments

We acknowledge experimental contributions from the whole Milano cryogenics group, and theoretical support from Prof. G. Benedek and Dr. A. Filipponi.

## References

[1] S.N. Ahmed et al., Phys. Rev. Lett. 92 (2004) 181301.
[2] J. Bonn et al., Nucl. Phys. B Proc. Suppl. 91 (2001) 273.
[3] A. Alessandrello et al., Physics Letters B 457 (1999) 253.
[4] M. Sisti et al., Nucl. Instr. Meth. A 520 (2004) 125.
[5] MARE, proposal submitted to INFN, Italy (2005). http://crio.mib.infn.it/wig/silicini/publications.html
[6] S.E. Koonin, Nature 354 (1991) 468.
[7] G. Benedek et al., Nucl. Instr. Meth. A 426 (1999) 147.
[8] F. Gatti et al., Nature 397 (1999) 137.
[9] C. Arnaboldi et al., submitted to Physical Reviev Letters.